\documentclass[prl,superscriptaddress, preprintnumbers,onecolumn]{revtex4-1} %,longbibliography %,showpacs
\usepackage{amsmath}
\usepackage{amssymb}
\usepackage{amsthm}
\usepackage{amsfonts}
\usepackage{algorithmic}
\usepackage{enumerate}
\usepackage{latexsym}
\usepackage{graphicx}
\usepackage{bm}
\usepackage{subfigure}
\usepackage{setspace}
\usepackage[dvips]{color}
\usepackage{hyperref}
\usepackage{bm}
\usepackage{mathrsfs}
\begin{document}

\title{Numerical Investigation of Localization in Two-Dimensional Quasiperiodic Mosaic Lattice}

\author{Hui-Hui Wang}
\affiliation{School of Physics and Materials Science, Guangzhou University, 510006 Guangzhou, China}
\affiliation{Huangpu Research and Graduate School of Guangzhou University, 510700 Guangzhou, China}

\author{Si-Si Wang}
\affiliation{School of Physics and Materials Science, Guangzhou University, 510006 Guangzhou, China}
\affiliation{School of Mathematics and Information Science, Guangzhou University, 510006 Guangzhou, China}

\author{Yan Yu}
\affiliation{SKLSM, Institute of Semiconductors, Chinese Academy of Sciences, P.O. Box 912, Beijing 100083, China}
\affiliation{School of Physical Sciences, University of Chinese Academy of Sciences, Beijing 100049, China}

\author{Biao Zhang}
\affiliation{School of Physics and Materials Science, Guangzhou University, 510006 Guangzhou, China}
\affiliation{Huangpu Research and Graduate School of Guangzhou University, 510700 Guangzhou, China}

\author{Yi-Ming Dai}
\affiliation{School of Physics and Materials Science, Guangzhou University, 510006 Guangzhou, China}

\author{Hao-Can Chen}
\affiliation{School of Physics and Materials Science, Guangzhou University, 510006 Guangzhou, China}

\author{Yi-Cai Zhang}
\affiliation{School of Physics and Materials Science, Guangzhou University, 510006 Guangzhou, China}

\author{Yan-Yang Zhang}
\email{yanyang@gzhu.edu.cn}
\affiliation{School of Physics and Materials Science, Guangzhou University, 510006 Guangzhou, China}
\affiliation{Huangpu Research and Graduate School of Guangzhou University, 510700 Guangzhou, China}
\affiliation{School of Mathematics and Information Science, Guangzhou University, 510006 Guangzhou, China}

\date{\today}

\begin{abstract}
A one-dimensional lattice model with mosaic quasiperiodic potential is found to exhibit interesting localization properties, e.g., clear mobility edges [Y. Wang et al., Phys. Rev. Lett. \textbf{125}, 196604 (2020)]. We generalize this mosaic quasiperiodic model to a two-dimensional version, and numerically investigate its localization properties: the phase diagram from the fractal dimension of the wavefunction, the statistical and scaling properties of the conductance. Compared with disordered systems, our model shares many common features but also exhibits some different characteristics in the same dimensionality and the same universality class. For example, the sharp peak at $g\sim 0$ of the critical distribution and the large $g$ limit of the universal scaling function $\beta$ resemble those behaviors of three-dimensional disordered systems.
\end{abstract}

%\pacs{}
\maketitle

\section{1. Introduction}
Anderson localization of wavefunctions in disordered systems is a subtle consequence of quantum interference\cite{Aderson1958,LocalizationReview2,LocalizationReview1,MarkosReview,Evers}. It is known that electrons in lower dimensions are more prone to localization in the presence of even weak disorder\cite{Abrahams79,LocalizationReview2,LocalizationReview1,Evers}. Theoretical progresses\cite{Theory02,Theory03,Theory04,Theory05,Theory06,Theory07} and experimental realizations\cite{Experiment01,Experiment02,Experiment03,Experiment04,Experiment05} of the delocalization-localization transition (or the metal-insulator transition, MIT) in disordered materials are still exciting topics until today. On the other hand, quasiperiodicity is a delicate structure between order and disorder\cite{QuasiCrystalRev1,QuasiCrystalRev2}.
The most typical case is the Aubry-Andr\'{e}-Harper (AAH) model, a one-dimensional (1D) lattice with nearest hopping and with a quasiperiodic potential incommensurate with the underlying lattice structure\cite{Harper,AA}. For this 1D model, there exist delocalized states at finite potential magnitude, which is impossible for a completely disordered counterpart\cite{Abrahams79,MacKinnon}. When the quasiperiodic potential magnitude is strong enough, all states will be localized.

Recently the study of quasiperiodic systems has attracted a lot of attentions due to inspiring progresses of experimental realizations\cite{QuasiCrystalRev1,QuasiCrystalRev2,Experiment2D,AAHExperiment1,AAHExperiment2,AAHExperiment3,AAHExperiment4}, and its relation with the many body localization\cite{AAHMBL}. New analytical methods based on Green's functions are developed, offering distinct insights into their physical grounds\cite{Analytical01,Analytical02}.
In the presence of pairing interaction, surprisingly, it was found that the quasiperiodic potential can enhance superconductivity remarkably\cite{EnhancedSC1,EnhancedSC2}, suggesting more unexpected phenomena of quasiperiodicity on quantum states. Even in the single particle picture, the quasiperiodic potential results in rich phenomena in different models.
Several nontrivial variations of the AAH model have been investigated recently, for instance, generalizations to a dimerized chain\cite{Roy} or two coupled chains\cite{CoupledChains}, with an unbounded quasiperiodic potential\cite{YiCaiZhang2022} or with a relative phase in the quasiperiodic potential\cite{AAH2023}, and higher dimensions\cite{AAModel3D}. Novel transport phases and topological phases are found when hoppings
are long ranged\cite{PowerLawHopping,LongRangeAAH,YWang} or quasiperiodic as well\cite{OffDiagonalMosaic}.
Besides these, two-dimensional (2D) quasiperiodic systems provide richer phenomena of localization\cite{MoireLocalization,Localization2D2020}, topology\cite{ShuChen2022}, flat band\cite{FlatBand2022}, and many-body effects\cite{Experiment2D,ManyBody2022}.

The original 1D AAH model possesses a self duality for the transformation
between the real and momentum spaces. This leads to the absence of mobility edges, which means that all eigenstates are either localized or delocalized, depending only on the strength of the potential. Among attempts to break the self duality, a nontrivial modification is the 1D quasiperiodic mosaic lattice, where the quasiperiodic potential only appears at sites with equally spaced distance\cite{WangY}. This is an analytically solvable model that exhibits mobility edges separating localized and delocalized states for a fixed potential strength.

In this manuscript, we generalize this quasiperiodic mosaic lattice to a two-dimensional (2D) version on the square lattice and study its localization properties. Different from its 1D counterpart, the phase boundary between localization and delocalization is highly fractal. By varying the initial phase of the quasiperiodic potential, statistical properties of the conductance are studied. In the end, we summarize all numerical results of conductance in a universal scaling function $\beta(g)$, which shows some different features from those in disordered 2D systems. Interestingly, for this 2D quasiperiodic model, some of these properties are similar to those in three-dimensional (3D) disordered systems.

\section{2. Model and Methods}

\begin{figure}[htbp]	
	\includegraphics*[width=0.33\textwidth]{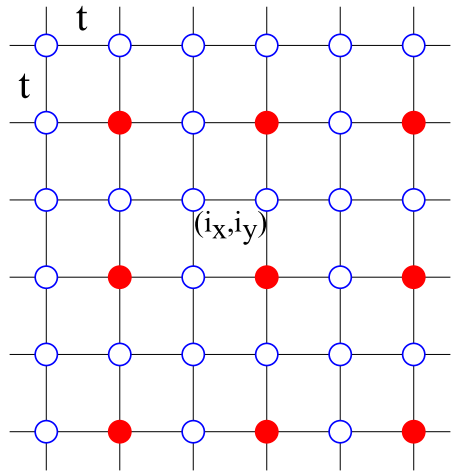}
	\caption{Illustration of our 2D quasiperiodic mosaic lattice model with $ \kappa=2$. Only solid red sites have nonzero quasiperiodic potential. Each site is represented by its coordinate $(i_x,i_y)$ and the nearest hopping is $t$. }
	\label{FigLattice}
\end{figure}
Our 2D model is defined on a square lattice with the nearest hopping Hamiltonian
\begin{equation}
\begin{aligned}
\mathcal{H} &= \sum_{i_x,i_y}t(c^{\dagger}_{i_x,i_y}c_{i_x+1,i_y}+c^{\dagger}_{i_x,i_y}c_{i_x,i_y+1})+\mathrm{H. c.} \\
	&+ 2\sum_{\bm{i_x,i_y}}V(i_x,i_y)c^{\dagger}_{i_x,i_y}c_{i_x,i_y},
\end{aligned}
\end{equation}
where $c^{\dagger}_{i_x,i_y}$ ($c_{i_x,i_y}$) creates (annihilate) an electron at the site with integral coordinate $(i_x,i_y)$.
In the following, the hopping integral $t=1$ will be used as the energy unit, and the lattice constant $a=1$ will be used as the length unit.

\begin{figure}[htbp]
    \includegraphics*[width=0.9\textwidth]{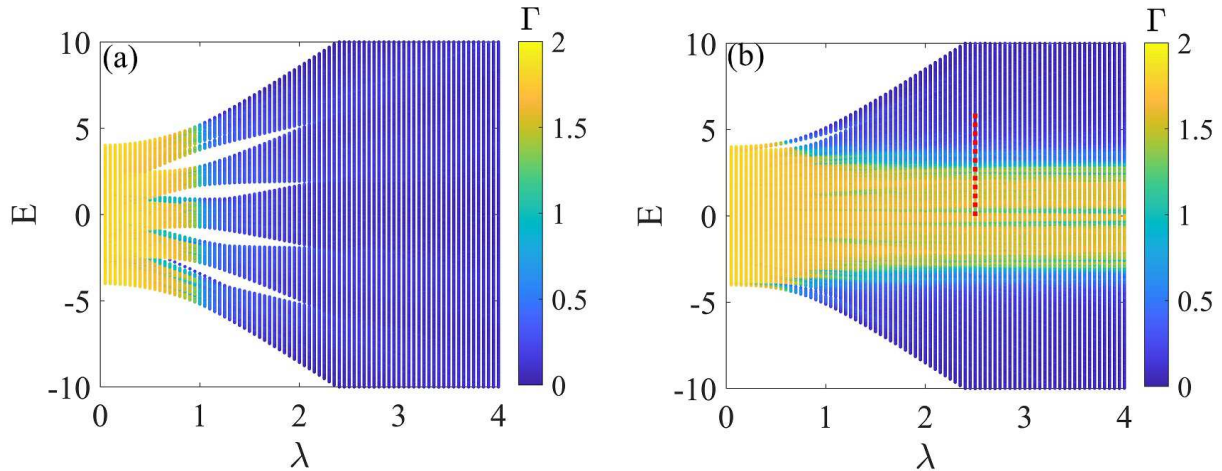}
	\caption{Fractal dimension $\Gamma$ of eigenstates as functions of corresponding eigenvalues $E$ and quasiperiodic potential strength $\lambda$, for a fixed sample size $L = 200$. (a) Non-mosaic with $\kappa=1$. (b) Mosaic with $\kappa=2$. }
	\label{FigGamma}
\end{figure}

The quasiperiodic mosaic potential $ V(i_x,i_y) $ is a generalization of the 1D version as\cite{WangY,Devakul}
\begin{subequations}\label{EqPotential}
\begin{align}
V(i_x,i_y) &=
\begin{cases}
 \lambda F(i_x,i_y),\enspace i_x=m\kappa \text{ and } i_y=n\kappa, \\
\qquad 0  \qquad,\qquad\text{otherwise}
\end{cases}\\
F(i_x,i_y)&=\cos\big[2\pi(\omega{i_x}+\theta_x)]+\cos[2\pi(\omega{i_y}+\theta_y)\big],
\end{align}
\end{subequations}
where $ \lambda\geqslant 0 $ is the amplitude of the potential. The irrational number $ \omega $ defines
the quasiperiodicity, and the integer $\kappa$ defines the mosaic period as illustrated in Figure \ref{FigLattice}. When $\kappa=1$ it returns to the original 2D AAH model without mosaic, and only models with $\kappa>1$ can be called mosaic. Recently an endoepitaxial growth of mosaic heterostructures has been experimentally realized on monolayer 2D atomic crystals\cite{Mosaic2DExp}.
In this manuscript, except in Figure \ref{FigGamma} (a), we always adopt $\omega=(\sqrt{5}-1)/2$ and $\kappa=2$,\cite{WangY} . The real numbers
$0\leqslant \theta_x,\theta_y<1$ are phase offsets of the potential profile. With other model parameters ($\lambda$, $\omega$ and $\kappa$) fixed,
different pairs of $(\theta_x,\theta_y)$ corresponds
to different realizations of an ensemble, similar to different realizations of a disordered ensemble\cite{Devakul}.
This will be useful when one needs statistical properties of this model, for example, mean values and statistical fluctuations.

Since analytical treatments for a 2D model are more difficult than those in 1D,
we will rely on numerical methods, which will be briefly introduced in the following.
All of our following calculations are performed in FORTRAN codes, along with Intel Math Kernel Library.

For a square shaped finite sample with $L\times L$ lattice sites,
the inverse participation ratio (IPR) of the $m$-th normalized eigenstate $|\Psi_{m,j}\rangle$ is defined as\cite{Evers}
\begin{equation}\label{EqIPR}
\text{IPR}(L,m)=\sum_{j=1}^{L\times L}|\Psi_{m,j}|^4.
\end{equation}
Then the localization of the state can be characterized by the fractal dimension of the wavefunction as
\begin{equation}\label{EqGamma}
\mathrm{\Gamma}(L,m)=-\frac{\ln\big[\text{IPR}(L,m)\big]}{\ln L}.
\end{equation}
In the thermodynamic limit $L\to \infty$, the state is extended (localized) if $\lim\limits_{L \to \infty}\Gamma(L) \to 2$ ($\lim\limits_{L \to \infty} \Gamma(L) \to 0$).

The localization properties can also be studied from quantum transports, by attaching two leads to the sample
with quasiperiodic potential. At zero temperature, the two-terminal conductance $G_T$ at Fermi energy $E$ is proportional to the transmission (Landauer formula)\cite{Landauer}, and
can be expressed as\cite{Datta,Datta2}
\begin{equation}
G_T(E)=2\frac{e^{2}}{h}g_T=2\frac{e^{2}}{h}\mathrm{Tr}\left[  \Gamma_{L}(E)G^{r}(E)\Gamma
_{R}(E)G^{a}(E)\right] , \label{EqLandauer}%
\end{equation}
where the prefactor 2 accounts for the spin degeneracy. Here
$G^{r/a}(E)\equiv \left(E\pm -H -\Sigma^{r/a}_L-\Sigma^{r/a}_R\right)^{-1}$ is the dressed retarded/advanced Green's function of the central sample, and $\Gamma_{L(R)}%
=i(\Sigma_{L(R)}^{r}-\Sigma_{L(R)}^{a})$ with $\Sigma_{L(R)}%
^{r/a}(E)$ being retarded/advanced self energies due to the left (right) lead, respectively.
In order to diminish the interface scattering, we take both leads to be lattices identical to the sample with vanishing potential $V$.

Based on the dimensionless conductance $g_T$ defined in Equation(\ref{EqLandauer}), the appropriate variable for size scaling is the intrinsic conductance $g$ expressed as \cite{Braun1997,Slevin2001}
\begin{equation}
\frac{1}{g}=\frac{1}{g_T}-\frac{1}{N_c},\label{Eqg}
\end{equation}
with $N_c$ the number of active channels at Fermi energy when the potential is absent.
The second term $\frac{1}{N_c}$ is used to deduct the effect of contact resistance so that the intrinsic transport property of the sample can be manifested. This intrinsic conductance of square shaped ($L \times L$) samples can be used to evaluate the scaling function $\beta =\frac{d \langle\ln g\rangle }{d \ln L}$ of the metal-insulator transition, where $\langle \cdots \rangle$ stands for averaging over an appropriate ensemble \cite{Slevin2001,LocalizationGraphene,XRWang2019}. An increase (decrease) of $ \ln g $ with increasing $\ln L$ indicates a metal (insulator) phase.

\section{3. Results}

\subsection{3.1. Phase diagram from eigenstates}

First let us have a global view of the localization property of this model. In Figure \ref{FigGamma}, we
present the fractal dimension $\Gamma$ of eigenstates as functions of potential magnitude $\lambda$ and corresponding eigenenergy $E$, for an isolated $200 \times 200$ sample with a definite potential configuration $\theta_x=\theta_y=0$. For comparison, we present the cases of $\kappa=1$ (non-mosaic) and $\kappa=2$ (mosaic) lattice in panels (a) and (b) respectively. With increasing $\lambda$, the band is broadened outwards and localized states ($\Gamma\sim 0$, blue color) appear. However, there is a remarkable difference between these two cases. For the non-mosaic case [Figure \ref{FigGamma} (a)], all states on the energy spectrum transit into localization simultaneously around $\lambda\sim 1$. In other words, there is no mobility edge for this case. For the mosaic case, on the other hand, there are both localized (blue color) and delocalized (orange color) states after $\lambda\gtrsim 0.6$.
This distinction between mosaic and non-mosaic models are similar to that in 1D \cite{WangY}. Also similar to the mosaic lattice in 1D\cite{WangY}, extended states ($\Gamma\sim 2$, yellow color) mostly distribute around the band center. Although their fraction among all states decreases with increasing $\lambda$, their existence can survive at large $\lambda$. In the following, we will focus on the mosaic lattice only.

\begin{figure*}[htbp]
    \includegraphics*[width=0.98\textwidth]{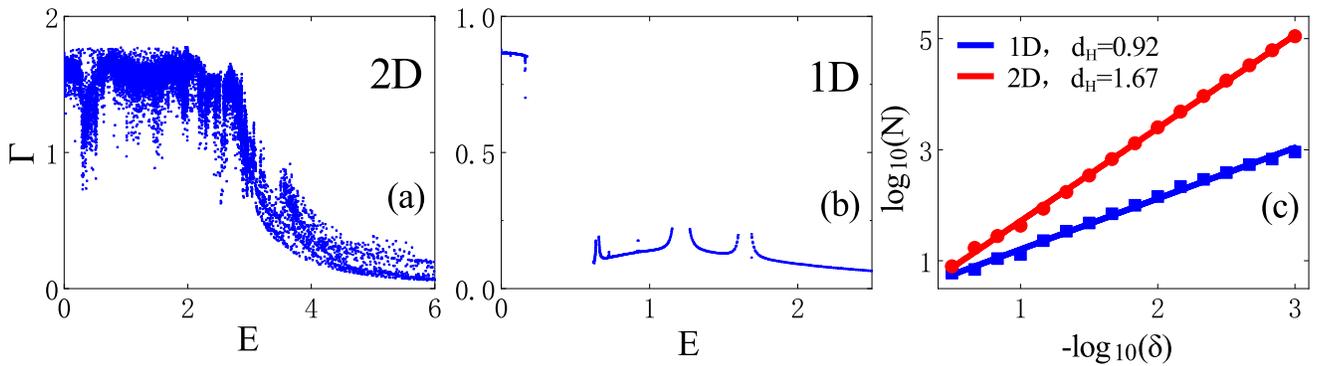}
 	\caption{(a) $\Gamma$ as a function of $E$ for $\lambda=2.5$, which is just the cross section shown as the red dotted line in Figure \ref{FigGamma} (b). (b) $\Gamma$ as a function of $E$ for a 1D mosaic lattice, with $\lambda=2.5$, $\kappa=2$ and $N_x=3000$. (c) The $\log - \log$ plots of box counting parameters $\delta$ versus $N$ for panel (a) (red) and panel (b) (blue), whose slope gives the Hausdorff dimension. }
 	\label{FigGammaProfile}
\end{figure*}

To see more details of this phase diagram in Figure \ref{FigGamma} (b) for the 2D mosaic lattice, we plot the profile of $\Gamma(E)$ along the red dashed line
at $\lambda=2.5$ in Figure \ref{FigGammaProfile} (a). For comparison, a typical counterpart of the 1D model is plotted in Figure \ref{FigGammaProfile} (b). Compared to the 1D case, the first obvious feature of the 2D model is strong fluctuations even around adjacent eigenenergies. For example, in the delocalization region $E\lesssim 3$, $\Gamma$ for most eigenstates fluctuate rapidly between $(1.35, 1.8)$, and some of them even drop around $\Gamma=1$. Furthermore, the transition between localization and delocalization tends to be a fractal region. In 1D, on the contrary, the profile consists of smooth curves with sudden jumps at transitions. Similar to that shown Figure \ref{FigGammaProfile} (a), fractal transitional region of transports is also predicted for the topological transition of a 2D incommensurate bilayer, which is a quasiperiodic structure as well\cite{YuYan2022}.

To characterize the difference between Figure \ref{FigGammaProfile} (a) and (b) quantitatively, we calculate the Hausdorff dimension $d_H$ of the point set $\big\{ (\Gamma_i,E_i) \big\}$ ($i$ is the index of eigenstates) shown in these 2D panels, by using the standard box-counting (BC) method \cite{Guarner,XYang}. This algorithm counts the number $N$ of squares with size $\delta$ which are necessary to continuously cover the graph of points $(\Gamma_i,E_i)$ rescaled to a unit square. In the intermediate region of $\delta$ (``the scaling region'') where the scaling relation $N \sim \delta^{-d_H}$ holds, the slope of the log-log plot of $\delta-N$ is the estimate of the Hausdorff dimension $d_H$ \cite{Veen,vanVeen}. The results corresponding to Figure \ref{FigGamma} (a) and (b) are shown as red and blue plots in panel (c) respectively. As expected, $d_H$ for the 1D model is close to $1$, reflecting the fact that $\Gamma(E)$ consists of simple smooth curves. For the 2D model however, $d_H\sim 5/3$ suggests that $\Gamma(E)$ is indeed highly fractal. We have checked that (but not shown here) for a very small potential magnitude, say,  $\lambda=0.3$, the relative fluctuation of $\Gamma(E)$ of the 2D model can be smaller, but its Hausdorff dimension ($d_H=1.68$) is still distinctly larger than that of the 1D model ($d_H=0.87$).

\subsection{3.2. Transport}
The above results were from eigenstates of an isolated sample. Now let us scrutinize transport properties
obtained by attaching conducting leads to the sample.

\begin{figure}[htbp]
	\includegraphics*[width=0.45\textwidth]{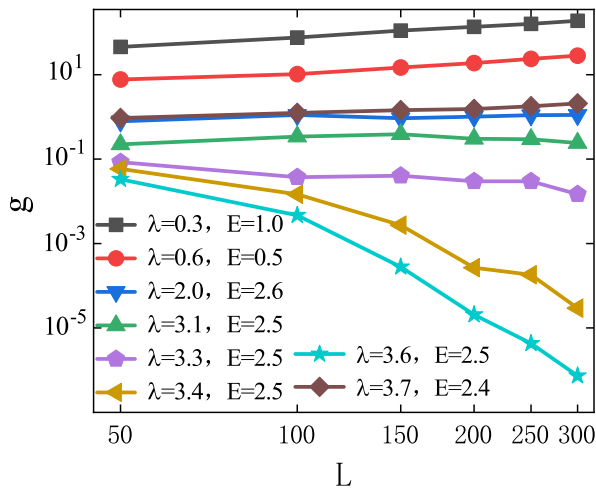}
	\caption{The typical value of the intrinsic conductance $\exp\big(\langle \ln g \rangle\big)$ as a function of sample size $L$, for different model parameters $\lambda$ and $E$. The sample is square shaped and the average $\langle \cdots \rangle$ is over 150,000 realizations of $(\theta_x, \theta_y)$. }
	\label{FigScaling}
\end{figure}

The localization (delocalization) can be characterized by the increasing (decreasing) of $g$ [Equation(\ref{Eqg})] with growing sample size $L$\cite{LocalizationReview1,Evers}. Similar to the case of disordered systems\cite{Slevin2001}, the conductance should be ensemble averaged before size scaling, to diminish the effect of sharp coherent fluctuations. To this end, we choose $\theta_x$ and $\theta_y$ to be random variables uniformly distributed within $(0,1)$, and define $\langle \cdots \rangle$ to be the arithmetic average over different realizations of $(\theta_x,\theta_y)$. In the following calculations, all averages are over 150,000 realizations. Since $g$ is not a self-averaging quantity and $\ln g$ is ``better distributed'' than $g$ itself (especially in the case of localization, as will be seen in the following), it is numerically more preferable to extract information from $\langle \ln g \rangle$ (or equivalently the typical value $g^{\text{typ}}\equiv \exp\big(\langle \ln g \rangle\big)$) rather than the mean value $\langle g \rangle$\cite{Slevin2001,Evers,MarkosReview}.
In Figure \ref{FigScaling}, we present the typical values of conductance $g^{\text{typ}}$
as functions of the sample size $L$, for different model parameters. There are typical delocalization and localization states with increasing or decreasing $g(L)$ dependence respectively, and also critical states ($\lambda=3.1$ and $E=2.5$) between them.

\begin{figure}[htbp]
	\includegraphics*[width=0.45\textwidth]{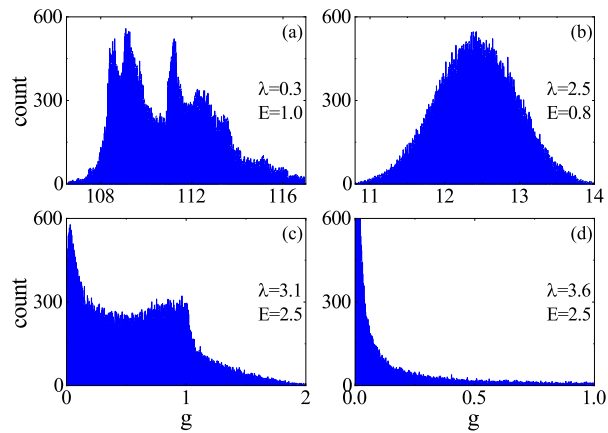}
	\caption{ Statistical distribution of $g$ with sample size $L=150$. (a) and (b): the delocalized states with $\beta>0$. (c) The critical state with $\beta=0$. (d) The localized state with $\beta<0$. The statistics is over 150,000 realizations of $(\theta_x,\theta_y)$.}
	\label{FigHistG}
\end{figure}

\begin{figure}[htbp]
	\includegraphics*[width=0.45\textwidth]{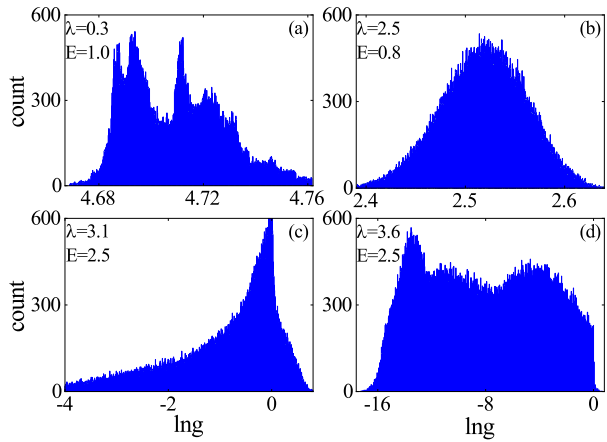}
	\caption{ Similar to Figure \ref{FigHistG} but for the quantity $\ln g$. }
	\label{FigHistlnG}
\end{figure}

Besides the averaged quantity, the distribution of the conductance also provides insightful information of localization\cite{Evers,MarkosReview}.
In Figs. \ref{FigHistG} and \ref{FigHistlnG}, we present distributions of $g$ and $\ln g$ respectively, for four typical regimes.
In Figure \ref{FigHistG} (a) with extremely weak disorder, the conductance has a very small relative deviation $\Delta g/ g$.
The distribution profile is rather irregular with multiple peaks. This is not surprising because the quasiperiodic potential is not ``random-like'' enough. Moreover, the energy scale associated with the quasiperiodic potential
has not dominated over the level separation of the finite sample yet,
so that the transmission can be very sensitive to the competition between these two factors, resulting in multiple peaks of the distribution.
With a stronger potential but still in the delocalized state as shown in Figure \ref{FigHistG} (b),
the distribution is smoothened to be a perfect Gaussian profile. This is identical to what happens in a delocalized
state with disorder\cite{MarkosReview}. From Figure \ref{FigHistlnG} (a) and (b) we can see that in the delocalized regime,
the distribution profile of $\ln g$ is almost identical to that of $g$.

At the critical state between localization and delocalization presented in Figure \ref{FigHistG} (c), the distribution profile is largely deformed.
There is an obvious nonanalytic point at $g=1$, which is also the common feature at the MIT in disordered systems\cite{Markos2DAndo,MarkosReview}. Another noticeable feature is the sharp peak near $g=0$. From the distribution of $\ln g$
shown in \ref{FigHistlnG} (c) it can be confirmed that this is a peak \emph{close to} $g=0$, instead of one \emph{at} $g=0$ \cite{Markos2DAndo}.
Interestingly, such a peak close to $g=0$ is also found at the MIT of a three-dimensional (3D) orthogonal system\cite{MarkosReview}, but is absent at that of a 2D symplectic system (the only example to exhibit bulk MIT in 2D)\cite{Markos2DAndo,Marko} and at the plateau-plateau transition of a 2D unitary system (quantum Hall effect)\cite{Distribution2D,Evers}.
In one word, for this 2D quasiperiodic model, the statistical distribution of the MIT   exhibits similar behavior to that of a 3D disordered system.

In the localized phase shown in Figure \ref{FigHistG} (d), the conductance displays a single-peak distribution
highly concentrated around $g=0$, which is also similar to that in the localized phase of disordered systems\cite{DistributionSoukoulis,DistributionSuslov}. In disordered systems with localization, it was found that the quantity $\ln g$ is
``better distributed'' as a partial Gaussian distribution terminated around $\ln g=0$\cite{DistributionSoukoulis,DistributionSuslov}.
The distribution of $\ln g$ for our model at strong localization is presented in Figure \ref{FigHistlnG} (d).
One can see that, although it is not a typical Gaussian shape but a clear termination around $\ln g$ still persists.

\subsection{3.3. Scaling function}
\begin{figure}[htbp]
	\includegraphics*[width=0.45\textwidth]{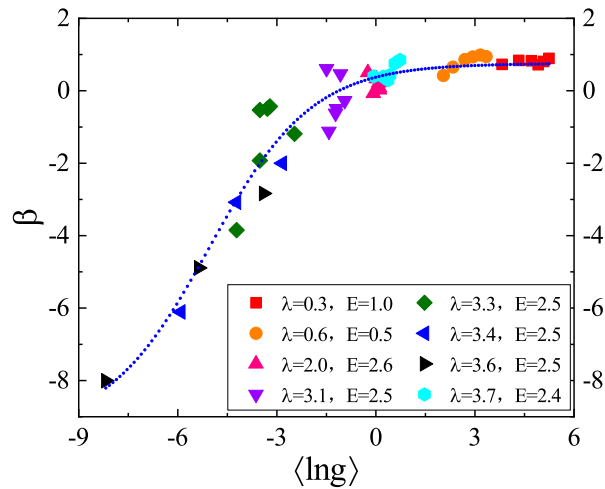}
	\caption{The scaling function $\beta(g)=d \langle \ln g \rangle / d \ln L$ as a function of $\langle \ln g \rangle$ in the two-dimensional quasiperiodic mosaic lattice.
Dots are numerical differential (see text) results from data shown in Figure \ref{FigScaling}. The dashed curve is a numerical fitting of these dots with the expression Equation(\ref{EqBetaFitting}). }
	\label{FigBeta}
\end{figure}

For a certain set of model parameters $\{ \lambda, \omega \}$, and after a polynomial fitting of data points $(\langle\ln g\rangle,\ln L)$ shown in Figure \ref{FigScaling}, one can obtain a numerical evaluation of the scaling function $\beta =\frac{d \langle\ln g\rangle }{d \ln L}$. Results of $\beta$ as a function of $\langle \ln g\rangle$ for a wide range of model parameters are shown in Figure \ref{FigBeta}. Within our best capability of calculation, i.e., 150,000 configurations at size $300 \times 300$, there are still strong fluctuations, especially around the transition region, $\beta \sim 0$, which is consistent with what we have seen from Figure \ref{FigGammaProfile} (a) and Figure \ref{FigHistG} (c). Nevertheless, some useful information can still be drawn. Firstly, all data points, similar to the case of disordered systems\cite{Slevin2001}, tend to collapse around a universal curve. We manage to fit these data with an appropriate nonlinear function. Numerically, among common functions we find that the Boltzmann function provides the best smooth curve that fits these data with the expression
\begin{eqnarray}
%\beta (\ln g)= -9.387-\frac{10.142}{ 1+\exp[(\ln g + 5.038)/1.550]
\beta (\ln g)&=& A_1+\frac{\displaystyle A_1-A_2}{\displaystyle 1+\exp\big[(\ln g - x_0)/\Delta \big] },\\ \label{EqBetaFitting}
A_1&=&-9.387,\qquad A_2=0.755,\nonumber \\
x_0&=&-5.038,\qquad \Delta=1.550 \nonumber
\end{eqnarray}
which is plotted as the blue dashed curve.
This is a monotonic function with a single zero point at $\ln g_c =-1.130$ where the MIT occurs. We stress again that this Boltzmann function is just an empirical choice instead of a principle-based analytical result, and we merely use it to extract some useful quantities in a numerically convenient way. For example, the slope of $\beta \big( \ln g \big)$ at its zero gives the inverse of $\nu$, the critical exponent characterizing the the divergence of the localization (correlation) length near the critical point\cite{LocalizationReview1,Slevin2001}. From Equation(\ref{EqBetaFitting}) we have an estimate $\nu \sim 2.22$. This value is larger than those of MITs in 3D, and that of MIT in 2D symplectic systems\cite{Slevin2001,DNShengCriticalExponents}, but is close to that of the plateau-plateau transition of 2D quantum Hall effects (unitary system)\cite{QHEnu1,QHEnu2,QHEnu3,QHEnu4}. Another important feature is the saturation of $\beta \sim 1$ at large $ \ln g $ limit. For disordered systems, it has been known that the $\epsilon$ expansion in the large conductance limit gives $\beta \rightarrow D-2$, for all three universality classes\cite{EpsilonExpansion1,EpsilonExpansion2,LocalizationReview2,Slevin2017}.
In this sense, the scaling behavior of our 2D quasiperiodic model in the weak potential (therefore large conductance) limit seems to resemble that of the 3D disordered model.

\section{4. Summary and Discussion}
In this paper, we numerically investigate the localization properties of the 2D quasiperiodic mosaic lattice model. We find some properties similar to, and also some different from those of 2D disordered systems.

Similar to the 1D counterpart, there exists localization-delocalization transition when varying the energy and the strength of the quasiperiodic potential. However the transition region is fractal like, contrary to clear phase boundaries in 1D. This model shares many common statistical features of the conductance $g$, for example, Gaussian-like distribution in the delocalized phase, high peak near zero in the localized phase, and the existence of non-analytical behavior of the distribution function at the critical point between two phases. However, this critical distribution at the small $g$ limit shows a sharp peak around zero, which is a particular feature of the 3D disordered systems. The scaling function $\beta$ is also a universal function of $g$, but its large $g$ limit approaches 1, which is also the feature of 3D disordered systems. This may suggest a novel role of the spatial dimensionality of quasiperiodic systems, which will be studied in the future. \cite{QuasiCrystalRev1,QuasiCrystalRev2}.

\section{Acknowledgements}
This work was supported by National Natural Science
Foundation of China under Grant Nos. 12104108, 11774336 and 11874127, the Joint Fund with Guangzhou Municipality under Nos. 202201020198 and 202201020137, and the Starting Research Fund from Guangzhou University under Grant Nos. RQ2020082, RQ 2020083 and 62104360.

\end{document}